\documentclass[useAMS,usenatbib,usegraphicx]{mn2e}

\usepackage{amssymb}
\usepackage{times}
\usepackage{subfigure}
\usepackage[T1]{fontenc}
\usepackage{aecompl}

\begin{document}

\title{On the Relative Ages of the $\alpha$-Rich and $\alpha$-Poor Stellar Populations in the Galactic Halo}
 \author[K. Hawkins et. al.]{K. Hawkins$^{1}$\thanks{E-mail: khawkins@ast.cam.ac.uk}, P. Jofr\'e $^{1}$, G. Gilmore$^{1}$, and T. Masseron$^{1}$  \\
$^{1}$Institute of Astronomy, Madingley Road, Cambridge. CB3 0HA}

\date{Accepted 2014 September 11.  Received 2014 September 10; in original form 2014 August 11}


\maketitle

\label{firstpage}

\begin{abstract}
We study the ages of $\alpha$-rich and $\alpha$-poor stars in the halo using a sample of F and G dwarfs from the Sloan Digital Sky Survey (SDSS). To separate stars based on [$\alpha$/Fe], we have developed a new semi-empirical spectral-index based method and applied it to the low-resolution, moderate signal-to-noise SDSS spectra. The method can be used to estimate the [$\alpha$/Fe] directly providing a new and widely applicable way to estimate [$\alpha$/Fe] from low-resolution spectra. We measured the main-sequence turnoff temperature and combined it with the metallicities and a set of isochrones to estimate the age of the $\alpha$-rich and $\alpha$-poor populations in our sample. We found all stars appear to be older than 8 Gyr confirming the idea that the Galactic halo was formed very early on. A bifurcation appears in the age-metallicity relation such that in the low metallicity regime the $\alpha$-rich and $\alpha$-poor populations are coeval while in the high metallicity regime the $\alpha$-rich population is older than the $\alpha$-poor population. Our results indicate the $\alpha$-rich halo population, which has shallow age-metallicity relation, was formed in a rapid event with high star formation, while the $\alpha$-poor stars were formed in an environment with a slower chemical evolution timescale. 
\end{abstract}


\section{Introduction}
We live in a typical spiral galaxy in the Universe, but what makes the Milky Way special is that only in our Galaxy can we probe the chemical, kinematical and spatial distribution of individual stars with high accuracy. Different patterns in these distributions are used to define different stellar populations and studying these in detail is crucial to understand how spiral galaxies form and evolve. In this context, the stellar halo is of paramount importance to reveal the early stages of the Milky Way because these stars are mostly very old and metal-poor \citep[e.g.][]{Mcwilliam1997, Unavane1996, Helmi2008, Jofre2011, Kalirai2012}.

After the classical work of \cite{Eggen1962}, where the stellar halo is composed of a simple stellar population which formed from the monolithic collapse of the proto-galactic cloud, many works have shown that the stellar halo is rather complex with components of stars being accreted after the initial collapse \citep{Searle1978, Ibata1994, Belokurov2006}. Much of the discussion in recent years has been focused on ways to quantify the amount of different stellar populations in a way to ascertain the importance of the merging history in the formation of the Milky Way. The works of \cite{Carollo2007, Carollo2010}, for example, have shown observational evidence that there is an outer stellar halo, which is dominant at galactocentric distances larger than 15 kpc, mostly formed from accreted sub-units and an inner stellar halo, confined to galactocentric distances less than 15 kpc, which seems to be rather smooth, with no predominant sub-unit like the outer halo. However, these results are still debated \citep[see discussion in][]{Ivezic2012, Schonrich2011}. 

Interestingly, inner halo stars seem to have two chemical patterns: a classical one with [$\alpha$/Fe] $\sim$ +0.40 which is related to the product of star formation in the large initial collapsing proto-galactic gas cloud (often thought of as the {\it in-situ} population); and another one with [$\alpha$/Fe] $\sim$ +0.20, which is related to the formation of stars in an environment of lower star formation rate, typically in smaller gas regions \citep[e.g.][hereafter N10]{Nissen2010}. The latter are attributed to have extra-galactic origins that were accreted onto the Milky Way after the formation of the main population (i.e. {\it accreted} stars). These two populations are commonly referred as ``$\alpha$-rich" and ``$\alpha$-poor" \citep{Nissen2010}, which have been subject of interest among other studies \citep{Ramirez2012, Sheffield2012}. These two populations have been extensively studied in a series of papers by Schuster \& Nissen. They have used a sample of high resolution and high signal-to-noise ratio (SNR) spectra of halo stars to show that these two populations are distinct in kinematics and abundances of $\alpha$-elements (N10), but indistinguishable from other chemical abundances such as Li and Mn \citep{Nissen2011, Nissen2012}. One interesting question arises with the findings of N10: do the $\alpha$-poor and $\alpha$-rich populations have the same age and/or is there correlation between age and metallicity? Determining the age differences between these two populations will help distinguish between the formation and assembly timescales of the Galactic halo. 

\cite{Schuster2012} attempted to answer this question by finding that the $\alpha$-poor stars in their sample are $\sim$ 2-3 Gyr younger than the $\alpha$-rich population. This is in favor of the models of \cite{Zolotov2009, Zolotov2010}. Yet, the recent simulations of \cite{Font2011} found that in situ stars can be as much as 3-4 Gyr younger than the accreted population. \cite{Schuster2012} argued that their $\alpha$-rich and $\alpha$-poor populations can be explained by a scenario where an initial disk/bulge formed in a monolithic collapse producing the $\alpha$-rich population. At later times, the $\alpha$-rich stars in the primeval disk were scattered into the halo via merging events that subsequently populated the $\alpha$-poor component of the halo. However, the conclusions of \cite{Schuster2012} were drawn from only 9 stars lying in the metallicity range of -0.4 $<$ [Fe/H] $<$ -1.40 dex. This made us wonder whether a difference in age between $\alpha$-poor and $\alpha$-rich can also be found using the data from the Sloan Digital Sky Survey \citep[SDSS,][]{York2000}, which contains thousands of halo stars extending the metallicity domain towards much lower metallicities compared to \cite{Schuster2012}. Although the spectra from SDSS have much lower resolution, it is still possible to rank the metal-poor stars in $\alpha$-abundance space. 

Most of the current methods developed for measuring $\alpha$-abundances with low resolution spectra attempt synthetic spectral matching \citep[e.g][]{Lee2011}. In these methods, a grid of synthetic spectra is constructed with relatively fine spacing in [$\alpha$/Fe] and degraded to low resolution. Other methods \citep[e.g.][]{Franchini2010, Franchini2011}, use a set of Lick indices to measure the [$\alpha$/Fe]. \cite{Franchini2011} determined the [$\alpha$/Fe] of a sample of F, G, and K stars observed with SDSS within $\pm$ 0.04 dex (accounting for the internal errors only). These methods heavily rely on having a good representation of real spectra through a synthetic grid of spectra and fairly well determined stellar parameters. The primary disadvantage to those methods, is that the estimated [$\alpha$/Fe] is strongly affected by the stellar parameter uncertainties because of the degeneracies that exist between the stellar parameters and the [$\alpha$/Fe]. These factors make estimating [$\alpha$/Fe] at low metallicity extremely difficult. We developed a new method which moves beyond the grid matching techniques, of current methods by defining an index, with the aid of a synthetic grid of spectra, which is computed using only the observed spectra. In this way our method is a semi-empirical way to estimate [$\alpha$/Fe]. Using this method, we can rank our sample based on their $\alpha$-abundance in a more model-independent way than the current methods. With this method, we aim to measure the age-metallicity relation of $\alpha$-rich and $\alpha$-poor halo field stars separately. 

Due to poor distance estimates, we did not attempt to determine ages with the standard isochrone fitting technique \citep[for a discussion on the method consult,][]{Soderblom2010} which was also employed by \cite{Schuster2012}. The large number of stars in our sample allows us to group stars in metallicity and determine the age of that population very accurately if there is a coeval dominant population \citep{Unavane1996, Schuster2006, Jofre2011}. This way of determining ages relies on using the colour (or temperature) of the main sequence turn-off of the population, which presents a sharp edge in the temperature distribution. By comparing the turn-off temperature of an $\alpha$-rich population at a certain metallicity and an $\alpha$-poor population of that metallicity, we can quantify the age difference between those populations. That age difference (if any), using a much larger sample of stars and a larger metallicity coverage, provides clues as to when the accreted stars of the inner halo formed relative to the in-situ ones.  That information is valuable for constraining theoretical models of the Milky Way formation. 

This paper is organized in the following way: In section \ref{sec:Sample} we define the sample of SDSS halo stars for which we estimated the $\alpha$-abundances and ages. In section \ref{sec:Method} we describe our new method to categorize low-resolution spectra based on their $\alpha$-abundances in the regime of the Galactic halo. In section \ref{sec:validation} we validate our method. In section \ref{sec:Ages} we employ the method to split our sample into a $\alpha$-rich and $\alpha$-poor population for which we determine the ages, and their errors in each population. In Section \ref{sec:discussion} we discuss the results and their implications for the formation of the Galactic halo. Finally, we summarize our findings in Section \ref{sec:summary}. 

\section{Data}
\label{sec:Sample}
This study made use of the SDSS/DR9 \citep{Ahn2012} and the SEGUE database \citep{Yanny2009}. SEGUE/SDSS provides approximately half a million low-resolution (R $\sim$ 2000) spectra which have stellar parameters estimates from the SEGUE stellar parameter pipeline \citep[SSPP,][]{Lee2008}. The spectra have wavelength coverage of 3900 -- 9000 \AA. We are interested in halo F- and G-type dwarfs, which allows us to determine the location of the turnoff. We selected these dwarf stars via a colour cut requiring the de-reddened $g-r$ colours to be in the range: 0.1 $<  (g-r)_{0} <$ 0.4 mag. We further required the SNR achieved to be at least 40 (see Section \ref{subsec:SNReffect}). To maximize the number of halo stars while deselecting other Galactic components we focused on the metallicity below -0.80 dex and further required the absolute Galactic latitude, $|$b$|$, to be larger than 30 degrees. While we expect some contamination from the thick disk in the most metal-rich bin, given the metallicity distribution function of \cite{Kordopatis2013b}, it is likely that our sample is dominated by halo stars. However, accurate distances and proper motions would be needed to fully resolve the space motion of the stars to study the contamination fraction.

\begin{figure}
\includegraphics[scale=0.43]{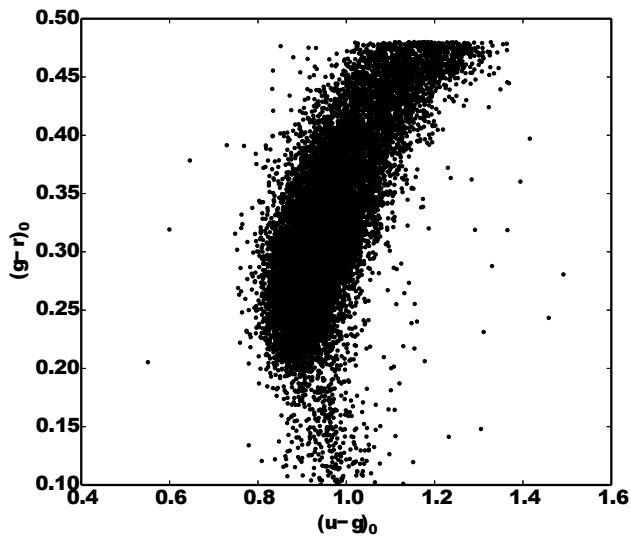}
\caption{The de-reddened colour-colour diagram of our F, G -type dwarf sample from SDSS.}
\label{fig:Colorcolor}
\end{figure}

We made additional cuts on the adopted SSPP parameters such that they are within the stellar parameter range defined in Section \ref{subsec:indexpreformance}. Finally, we co-added any duplicated spectra to increase the SNR. Our final sample contains 14757 unique objects. A colour-colour diagram of the sample can be found in Figure \ref{fig:Colorcolor}. The $(g-r)_{0}$ colour is directly related to the temperature of the star. Thus, Figure \ref{fig:Colorcolor} illustrates the main-sequence F and G dwarf stars of our sample are plentiful until $(g-r)_{0} \sim$ 0.2 mag which is approximate location of the turnoff. Above this temperature, i.e. smaller $(g-r)_{0}$ colours, are likely blue stragglers. Given the SNR cut and the fact that we select dwarfs; our sample is limited to stars near the sun (likely within 1-2 kpc) and thus the inner halo. 

\section{A Method to Estimate $\alpha$-Abundances}
\label{sec:Method}

\subsection{Grid of Synthetic Spectra}
\label{subsec:syntheticgrid}
We used a grid of synthetic spectra to develop and test our spectral-index method. In the grid, $\alpha$-enhancement is achieved by increasing (or decreasing) in lockstep the individual $\alpha$-elements (Ca, Ti, Si, Mg, O) from their solar values. The synthetic spectra make use of the 1D LTE MARCS model atmospheres of \cite{Gustafsson2008} which have a variety of $\alpha$-abundances. The synthetic grid was created using the Turbospectrum synthesis code \citep{Alvarez1998, Plez2012} which uses the line broadening treatment described by \cite{Barklem1998}. Solar abundances were taken from \cite{Asplund2005}. Atomic lines used by Turbospectrum are sourced from VALD, \cite{Kupka1999}, \cite{Hill2002}, and \cite{Masseron2006}. Line-lists for the molecular species are provided for CH (T. Masseron et al. 2014, in press), and CN, NH, OH, MgH and C2 (T. Masseron, in prep); the lines of SiH molecules are adopted from the Kurucz linelists and those from TiO, ZrO, FeH, CaH from B. Plez  (private communication). Microturbulence velocity for each spectrum were estimated using a polynomial relationship (Bergemann et al., in preparation) between microturbulence velocity and surface gravity developed for the Gaia-ESO Survey \citep{Gilmore2012}. The final grid covers 3000 K $\leq$ Teff $\leq$ 8000 K in steps of 200 K, 0.0 $\leq$ log g $\leq$ 5.0 in steps of 0.2 dex, and -3.0 $\leq$ [Fe/H] $\leq$ +1.0 in steps of 0.1 dex and -0.1 $\leq$ [$\alpha$/Fe] $\leq$ +0.4 dex in steps of 0.1 dex. The synthetic grid has only been used to provide a starting point to inform our placement of spectral bands which are sensitive to [$\alpha$/Fe].

\subsection{Spectral-Index Method (SIM)}
\label{subsec:SIM}
Our aim is to classify stars as $\alpha$-rich or $\alpha$-poor on the basis of an index that is relatively insensitive to stellar parameters. Following similar methods, often employed when studying stellar populations in other galaxies \citep[e.g.][]{Thomas2003}, we used a spectral index that is very sensitive to $\alpha$-abundances. This allows us to use a simplistic approach to find the $\alpha$-abundances using low-resolution SDSS spectra. The index was built with a combination of (1) spectral bands that are sensitive to $\alpha$-abundances and (2) control bands whose response to the stellar parameters mimic the response function of the $\alpha$-sensitive bands. All spectral bands must be large enough as to not be dominated by noise, yet not too large to be dominated by broad structures in the spectrum. Since we are interested primarily in [$\alpha$/Fe], we only choose bands which are sensitive to $\alpha$-abundances to be the main driver of our index diagnostic. We employed a semi-automatic method to explore the synthetic spectral grid to find moderate-size (15 \AA $<$ band width $<$ 80 \AA) spectral bands. We found five bands that are sensitive to the $\alpha$-abundances (listed in the top part of Table \ref{tab:bands}). We combined these bands in a linear way to obtain an index sensitive to [$\alpha$/Fe]. 

\begin{figure*}
\includegraphics[width=2.1\columnwidth]{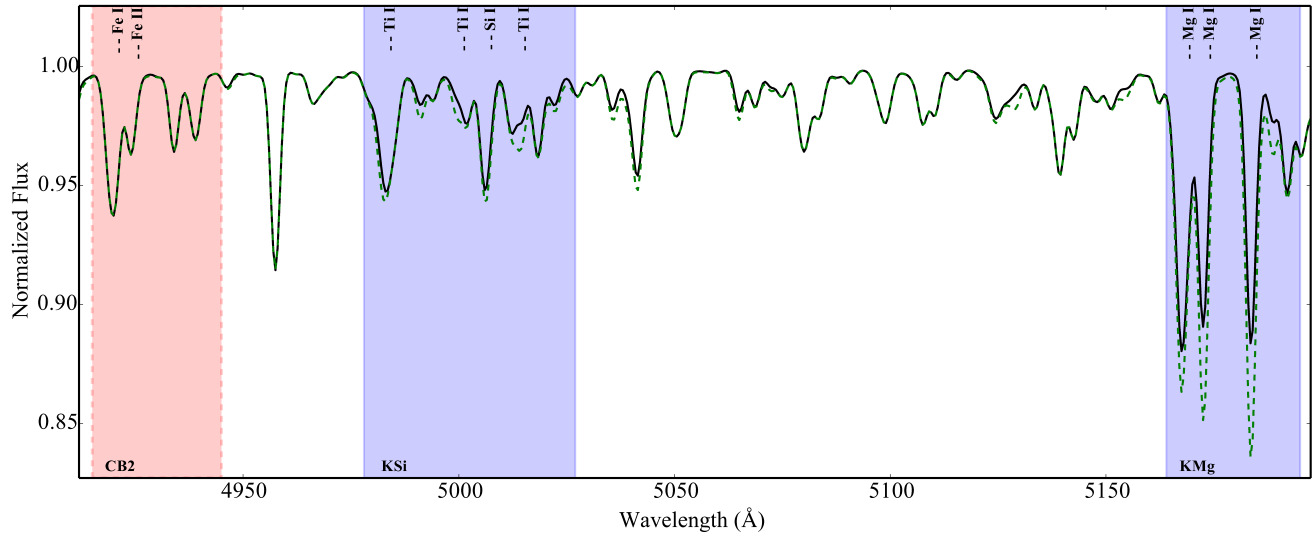}
\caption{Illustration of the KSi and Mgb (blue shaded regions), CB1 (red shaded region) spectral bands for an $\alpha$-rich (green dotted) and $\alpha$-poor (black solid) spectra with Teff = 6200 K, log g = 4.4 dex, and [M/H] = -1.50 dex.}
\label{fig:spectralex}
\end{figure*}

The control bands, on the other hand, were found via an iterative process by systematically scanning a grid of possible control bands. A good control band is one that minimizes the scatter in the index over a range of stellar parameters while maximizing the mean difference between the index at [$\alpha$/Fe] = 0.0 dex and [$\alpha$/Fe]  = +0.4 dex. We found three control bands, whose wavelength ranges are listed at the bottom of Table \ref{tab:bands}. Figure \ref{fig:spectralex} shows two synthetic spectra, one of with solar-scaled $\alpha$-abundances (black solid line) and the other that is $\alpha$-enhanced (green dotted line). The spectral regions are designated with red dashed lines corresponding to control bands and blue solid lines for $\alpha$-sensitive bands. 

With the set of spectral bands defined in Table \ref{tab:bands}, we define our $\alpha$ index diagnostic by:
\begin{equation}
\label{eq:index}
\textrm{Index} = \frac{\textrm{5Mgb+4KTi+3KSi+KCa+K1}}{\textrm{CB1+CB2+CB3}} .
\end{equation}
The weights in the equation are motivated by the line-strengths of the $\alpha$-sensitive bands, and were determined after trying several combinations. The proposed method differs from current methods which rely on a synthetic grid to model out the effects of the stellar parameters  \citep[e.g.][]{Thomas2003,Franchini2010, Franchini2011}. The advantage of our method is that we can use the spectrum alone to estimate the [$\alpha$/Fe] with relatively little effect from the uncertainties in the stellar parameters provided they are within the stellar parameter range defined in Section \ref{subsec:indexpreformance} below. 

From a physical point of view, the numerator of the index is defined by spectral bands that are centred near the spectral features of the $\alpha$-elements. Therefore, an increase in the $\alpha$-abundance causes an increase in the strength of the $\alpha$-sensitive spectral bands. An example of this can be seen in the KSi spectral band in Figure \ref{fig:spectralex} which is centred on the a series of Ti and Si features. The control bands (e.g. CB1) in the denominator are centred near Fe-peak elements (mostly Fe spectral features) yet have very similar transition properties (e.g. oscillator strength and excitation potential) as the $\alpha$-sensitive bands. As a result of this, the index tracks $\alpha$-abundances divided by Fe-peak abundances while simultaneously controlling for the stellar parameters. We note that the CB2 and KCa bands overlap. While the Fe and Cr lines in the CB2 band overlap in wavelength space with the KCa band, the response of the CB2 band to the stellar parameters is different than the $\alpha$-sensitive lines in the KCa band. Tests have shown that excluding the overlapping KCa band region from the CB2 band increases the scatter in the index at high $\alpha$-abundance by a factor of 1.4 leading to a less precise estimate. 

 \begin{table}
\caption{Spectral Bands defined in the Index.} 
\centering 
\begin{tabular}{c c c} 
\hline\hline 
Band&$\lambda_{i}$&$\lambda_{f}$ \\
&(\AA)&(\AA)\\
\hline 
KTi&4510&4591 \\
KSi&4978&5027 \\
Mgb&5164&5195 \\
KCa&5258&5274 \\
K1&6148&6172 \\
\hline
CB1&4915&4945 \\
CB2&5225&5275 \\
CB3&5390&5430 \\

\hline 
\end{tabular} 
\label{tab:bands}
\end{table}

\subsection{Processing Spectra}
\label{subsec:processing}
Synthetic and object spectra were preprocessed before searching for and measuring spectral features. All spectra were preprocessed in the same way so that it is possible to compare them directly. We applied this method to the SDSS spectra and as such their spectral parameters (e.g. resolution, etc.) defined the preprocessing parameters. We started by converting the wavelength scale of all spectra to an air-based system. Synthetic, and validation spectra were degraded to a resolution of 2,000 using a Gaussian kernel. All spectra were resampled to 0.5 \AA~ pixels over a wavelength range of 4000 -- 7000 \AA. 

We defined the local continuum via a boosted-median pseudo-continuum \citep[BMC,][]{Rogers2010}. This method is defined by taking a window around any particular wavelength and setting the continuum to be at a prescribed percentile of the fluxes within the window. The BMC method is based on two parameters: (1) the width of the window around a given wavelength and (2) the percentile of the flux which defines the pseudo-continuum. We started by assuming the parameters used in \cite{Rogers2010}, a window of 100 \AA~with a percentile of 90 per cent. The effect of the choice of these parameters is analyzed in detail in Section \ref{subsec:continuum}.

\subsection{Performance of the Index on Synthetic Spectra}
\label{subsec:indexpreformance}
We are interested primarily in F, G, and K dwarf stars and thus we restricted the stellar parameters to:
\begin{itemize}
\item 5000 K $\leq$ Teff $\leq$ 7000 K
\item log g $\geq$ 3.5 dex
\item -2.0 dex $\leq$ [M/H] $\leq$ 0.0 dex
\end{itemize}

Selecting warm stars means that we avoided significant molecular features that would require a more careful calibration in the index. Therefore, we required that the effective temperature must be at least 5000 K. However, if the temperature is too high, the lines we need for our diagnostic will be weaker and thus we set an upper limit in temperature at 7000 K, potentially excluding any very young metal-poor stars. We required the metallicities to be above -2.0 dex because below that limit it can be difficult, especially at high temperatures, to estimate the $\alpha$-abundance given the lack of strong lines (for more discussion see Section \ref{subsec:SPeffect}). The log g cut is to ensure we have selected dwarf stars. It is interesting to note that this simple method is used to search for spectral peculiarities, in our case stars with strong features around the $\alpha$-elements. However, the method could, in theory, be re-calibrated to  search for other scientifically exciting targets such as s-process or r-process enriched stars or others. This method could also be adapted to determine the abundance of other chemical species using low-resolution spectra.

\begin{figure*}
\includegraphics[scale=0.45]{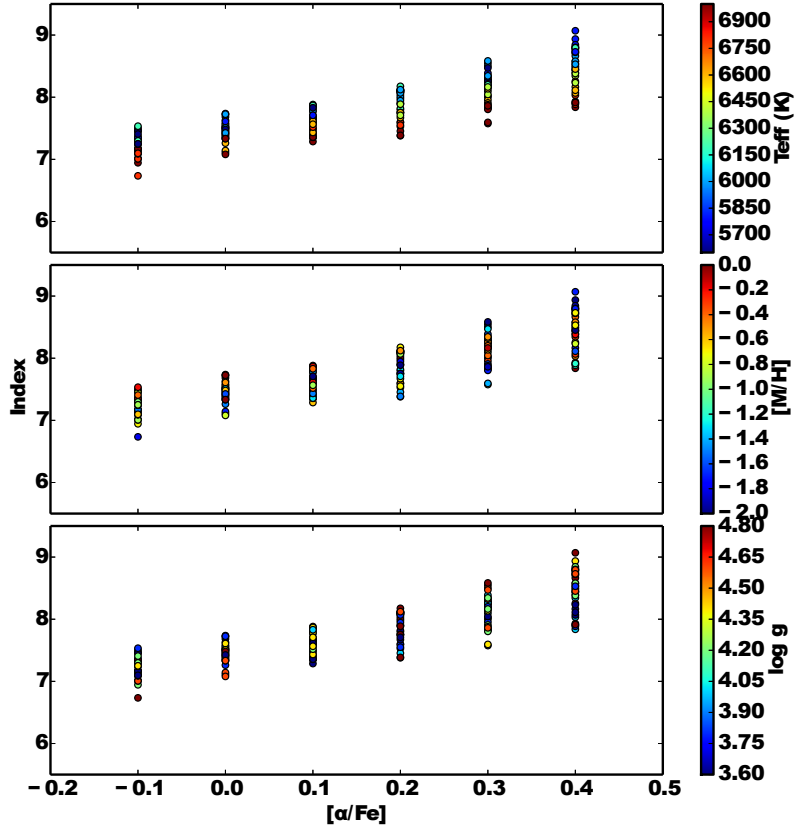}
\caption{The index value as a function of [$\alpha$/Fe]  colour coded by Teff (top panel), [Fe/H] (middle panel), log g (bottom panel) for a sample of 200 synthetic spectra.}
\label{fig:Tspecindex}
\end{figure*}

In Figure \ref{fig:Tspecindex} we plot the value of the index for a sample of 200 synthetic stars (selected from the parameter space and synthetic grid above), where the y-axis labels the index value and the x-axis is the synthetic value for [$\alpha$/Fe]. The colour code represents the temperature, metallicity, and surface gravity in the top, middle, and lower panels, respectively. We see a tight, linear correlation between the index value and the [$\alpha$/Fe], such that high index values indicate high $\alpha$-abundances. There does not appear to be any major systematic variations as a result of the stellar parameters. Figure \ref{fig:Tspecindex} shows that it is possible to distinguish between an $\alpha$-rich and $\alpha$-poor population as the mean index of stars with [$\alpha$/Fe] $\leq$ +0.25 dex is smaller than the mean index of stars with [$\alpha$/Fe] $>$ +0.25 dex. A star with an index below approximately 7.4 can be interpreted as $\alpha$-poor while a star with an index about 7.6 $\alpha$-rich. Figure \ref{fig:Tspecindex} also suggests that it may be possible to calibrate the index in a linear way to estimate the [$\alpha$/Fe] directly (which we explore in Section \ref{subsec:convertindex}).

\subsection{Effects of Stellar Parameters}
\label{subsec:SPeffect}
\begin{figure}
\includegraphics[width=\columnwidth]{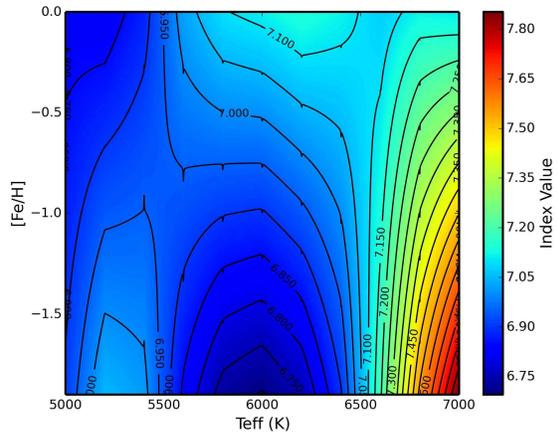}
\caption{The value of the index (colour) as a function of both the effective temperature and the metallicity for all dwarfs in the synthetic grid at [$\alpha$/Fe] = 0.00 dex}
\label{fig:SPeffect}
\end{figure}

In this section, we study the effect of the stellar parameters on the index. The index is a linear combination of $\alpha$-sensitive spectral bands divided by a set of control spectral bands. Ideally, the index should be constant with a given [$\alpha$/Fe] regardless of the stellar parameters. In practice, the stellar parameters have a small effect on the index. To study this effect, we computed the index over a broad range of stellar parameters (outlined in Section \ref{subsec:indexpreformance}) at a constant [$\alpha$/Fe]. Figure \ref{fig:SPeffect} shows the value of the index (color) as a function of both temperature and metallicity. The plot illustrates that the index is stable and only varies on the order of 1-2 per cent at effective temperatures below $\sim$6500 K. However, the index seems to vary significantly at the very metal-poor and high temperature end (bottom right corner of Figure \ref{fig:SPeffect}). This is expected as the metal-poor hot stars will likely have significantly weaker lines. From this figure, we can see that the method does not preform nearly as well for the hot (particularly at temperatures above 7000 K), metal-poor stars. 

\subsection{Effects of Signal-to-Noise}
\label{subsec:SNReffect}
\begin{figure}
\includegraphics[scale=0.45]{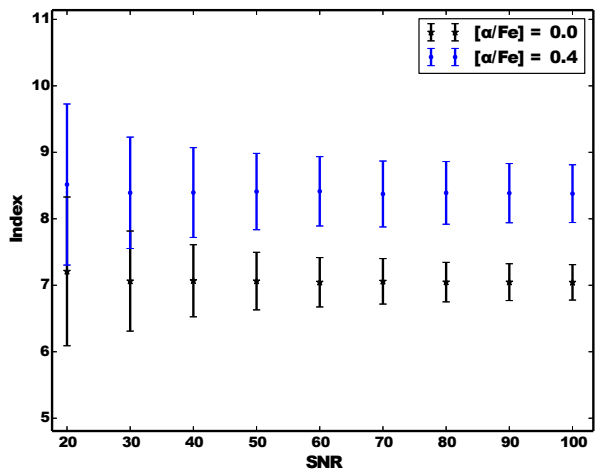}
\caption{The mean and spread in the index over the full range in stellar parameters as a function of SNR.}
\label{fig:SNReffect}
\end{figure}

To study the effect of both the signal-to-noise ratio (SNR) and the stellar parameters on the stability of the index, we have completed noise-injection experiments. We used the synthetic grid of spectra described in Section \ref{subsec:syntheticgrid}. Once these spectra were preprocessed, we added Gaussian white noise to simulate spectra at SNR between 20 and 100 typical for SDSS data. We then preformed a BMC continuum normalization. Finally, we calculated the index and plotted the variation in the index as a result of the stellar parameters at each SNR for [$\alpha$/Fe] = 0.00 dex and [$\alpha$/Fe] = +0.40 dex. Both the SNR and stellar parameters cause a variation in the index at a constant [$\alpha$/Fe]. 

Figure \ref{fig:SNReffect} shows that it is possible to separate [$\alpha$/Fe] = 0.00 and [$\alpha$/Fe] = +0.40 dex at SNR larger than 40. We find, at a SNR = 40, the total internal error on the index, which is defined as its mean variation, is found to be 0.60. Figure \ref{fig:SPeffect} illustrates that this value can be decreased significantly by further restricting the parameter space below 6500 K. For example, by restricting the  the parameter space to metal-poor halo turnoff stars (6000 $<$ T$_{\mathrm{eff}} <$ 6500~K, log g $>$ 3.5~dex, and -2.0 $<$ [Fe/H] $<$ - 0.80~dex), the total internal error 0.20 instead of 0.60.

\subsection{Effect of Pseudo-Continuum Placement}
\label{subsec:continuum}
We used the BMC method to determine the local continuum for our spectra that has the benefit of better control the continuum in the local region of our bands. It also has the advantage of reducing the uncertainty in the total flux of our measured spectral bands \citep{Rogers2010}. The BMC method is sensitive to two parameters and a balancing act must be played with both. These parameters are:
\begin{itemize}
\item The width of window around any spectral band - The width must be large in order to avoid small scale fluctuations and noise in the spectra but not too large as to encode large-scale structure (e.g. G band at $\sim$ 4300~\AA) 
\item The percentile of flux within the window which is defined as the continuum - The percentile must be high enough to 'see' the true continuum but cannot be so high that it becomes only sensitive to noise
\end{itemize}

\begin{figure}     
      \includegraphics[width=\columnwidth]{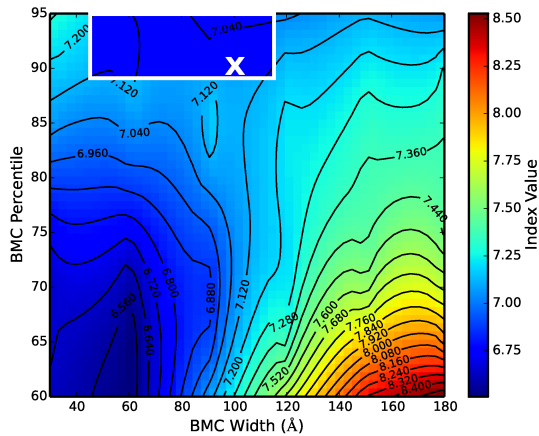}
       \caption{Stability of index as a function of BMC parameters. 'X' is the location suggested by Rogers et al. (2010). The shaded region represents the region where the index varies on an order of less than 1 per cent.}
        \label{fig:BMCsen}
\end{figure}

Therefore, it is useful to study the sensitivity of our index to these two parameters that define the placement of the continuum. To do this we took a single random synthetic spectrum and plotted the contours of the measured index for that spectrum as a function of both the width and percentile. The index should be insensitive to our choice of BMC parameters and thus we searched for regions where the derivative of the index to the BMC parameters is minimized. Remembering that the density of contours is proportional to the derivative of the index as a function of the BMC parameters, we selected a region that the index is very stable over a range of BMC parameters. The choice of BMC parameters adopted from \cite{Rogers2010}, which is the 'X' in Figure \ref{fig:BMCsen}, worked very well because the index does not seem to depend strongly on the choice of BMC parameters. This experiment has shown that as long as the percentile is between 89 and 95 per cent and the width is between 45 and 115 \AA~the index varies by less than 1 per cent. These results are independent of our choice of test spectra. We therefore define the pseudo-continuum at any wavelength as the 90th percentile flux in 100 \AA~window around that wavelength. 

\section{Validation}
\label{sec:validation}
To achieve a good validation we needed several validation sets. The various datasets were employed to allow us to study the performance of our method in different regimes of stellar parameter space, across large datasets, and with low-resolution SDSS data. We describe the validation sets and the reason they were chosen separately below.

\subsection{Comparison with the ELODIE Library}
The ELODIE library is a publicly available library of about 2,000 spectra of 1,388 stars observed with the ELODIE spectrograph on the Observatoire de Haute-Provence 1.93-m telescope. ELODIE3.1 \citep{Prugniel2001, Moultaka2004} contains high-resolution spectra (R = 42,000), high SNR ($>$ 100) observations of stars with 3700 K $<$  Teff $<$ 13600 K, 0.03 dex $<$ log g $<$ 5.86 dex and -2.8 dex $<$ [Fe/H] $<$ +0.17 dex. \cite{Lee2011} used a sample of 425 ELODIE spectra that have well known stellar parameters and their [$\alpha$/Fe] from the literature. It is important to note that because the [$\alpha$/Fe] values are taken from the literature there may be some scatter that exist in this validation set \citep{Lee2011}. 

\begin{figure*}
\includegraphics[scale=0.45]{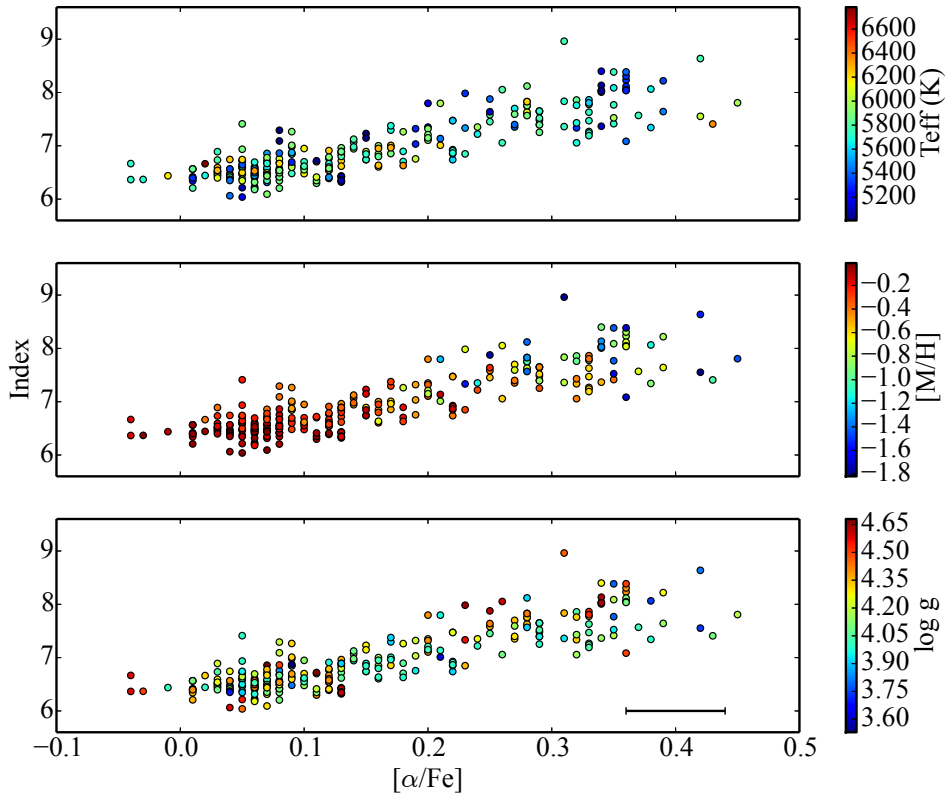}
\caption{The index value as a function of [$\alpha$/Fe] colour coded by Teff (top panel), [Fe/H] (middle panel), log g (bottom panel) for F, G, and K metal-poor dwarf ELODIE stars that are within our parameter range. The error bar represents the typical errors on the [$\alpha$/Fe] from the ELODIE high-resolution validation set.}
\label{fig:ELODIEindex}
\end{figure*}

We processed the ELODIE spectra in the same way as the rest of the spectra: degrading the resolution to R = 2,000 using a Gaussian kernel, 
resampling the spectra to 0.5 \AA~pixels, applying a BMC normalization. We calculated the index for all stars that fall in our parameter range and studied the index as a function of [$\alpha$/Fe] (Figure \ref{fig:ELODIEindex}). The colour scheme is the same as Figure \ref{fig:Tspecindex}. 

We see a tight correlation between the index and the $\alpha$-abundance over the full range of stellar parameters indicating that it is possible to estimate the [$\alpha$/Fe] directly from the index. This result shows us that we can rank stars from low to high [$\alpha$/Fe] using the index only. A star with an index below approximately 7.2 would be considered $\alpha$-poor while a star with an index about 7.4, $\alpha$-rich. These index values are consistent with the synthetic spectra in Section \ref{subsec:indexpreformance}. Although a large number of stars exist in the ELODIE library, they are mostly metal-rich compared to the halo stars we are interested in. As such, this validation set is not completely suitable for our purposes. Thus we must explore other validation sets.

\subsection{Comparison with Nissen \& Schuster Data}
\begin{figure*}
\includegraphics[scale=0.40]{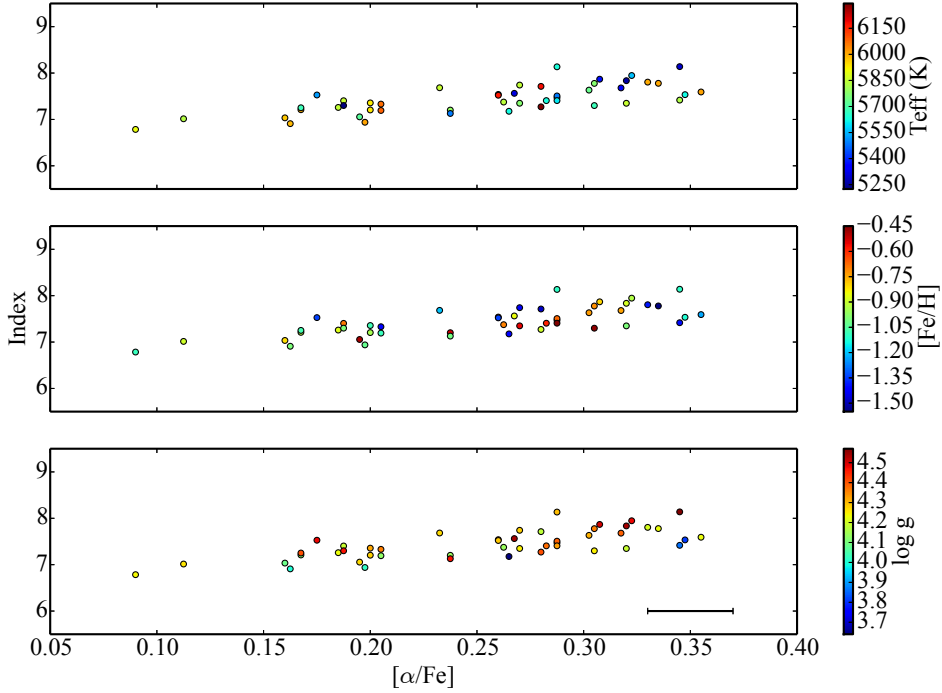}
\caption{The index value as a function of [$\alpha$/Fe] colour coded by Teff (top panel), [Fe/H] (middle panel), log g (bottom panel) for the NOT/FIES subset of the N10 data. The error bar represents the typical errors on the [$\alpha$/Fe] from the N10 high-resolution validation set.}
\label{fig:Nissenindex}
\end{figure*}

 \cite{Nissen2010} measured the Mg, Ti, Si, Ca, and Fe-peak elemental abundance abundances of 78 halo stars and 16 disk stars using spectra from the Very Large Telescope's (VLT) Ultraviolet and Visual Echelle Spectrograph (UVES) spectra and the Fiber fed Echelle Spectrograph (FIES) on the Nordic Optical Telescope (NOT). These spectra were kindly provided to us by P. Nissen. The wavelength coverage of VLT/UVES sample extends only to $\sim$ 4700 \AA. Since the index includes a Ti I feature at $\sim$ 4500 \AA, the VLT/UVES data from N10 could not be used for validation. The NOT/FIES spectra have a wavelength coverage that allows us to use this Ti I feature. In total, there are 47 stars from the N10 dataset that meet our criteria in wavelength coverage and stellar parameters space. N10 represents the parameters space we are most interested in, and is also internally consistent. This dataset is rather small and thus this set alone does not allow detailed analysis of the performance of the index but compliments the ELODIE validation set. Figure \ref{fig:Nissenindex} shows the performance of the index as a function of [$\alpha$/Fe] for the N10 data. The colour coding represents the stellar parameters as in Figure \ref{fig:Tspecindex}.

We found again a tight correlation between the index and the $\alpha$-abundance over the full range of stellar parameters. Figure \ref{fig:Nissenindex} makes it clear that we can use the index alone to categorize the high-$\alpha$' and 'low-$\alpha$' stars from N10. Stars with an index below 7.2 would be considered $\alpha$-poor while a star with an index of 7.4 is $\alpha$-rich. These values are the same as for the synthetic grid and the ELODIE validation sets. The tightness of the linear correlation between the index and the [$\alpha$/Fe] is further proof that we may be able to estimate the $\alpha$-abundance directly from the index.

\subsection{Comparison with SDSS Calibration targets}
Finally we used a sample of SDSS spectra with accurate stellar parameters and [$\alpha$/Fe] determined from high-resolution spectra. This sample was used in \cite{Lee2011} and was originally constructed to validate the SSPP pipeline \citep[for more details consult][]{AllendePrieto2008,Smolinski2011}. We used a subset, totaling 73 stars, from the SDSS calibration targets discussed in \cite{Lee2011} that exist inside our stellar parameter range. 

\begin{figure*}
\includegraphics[scale=0.43]{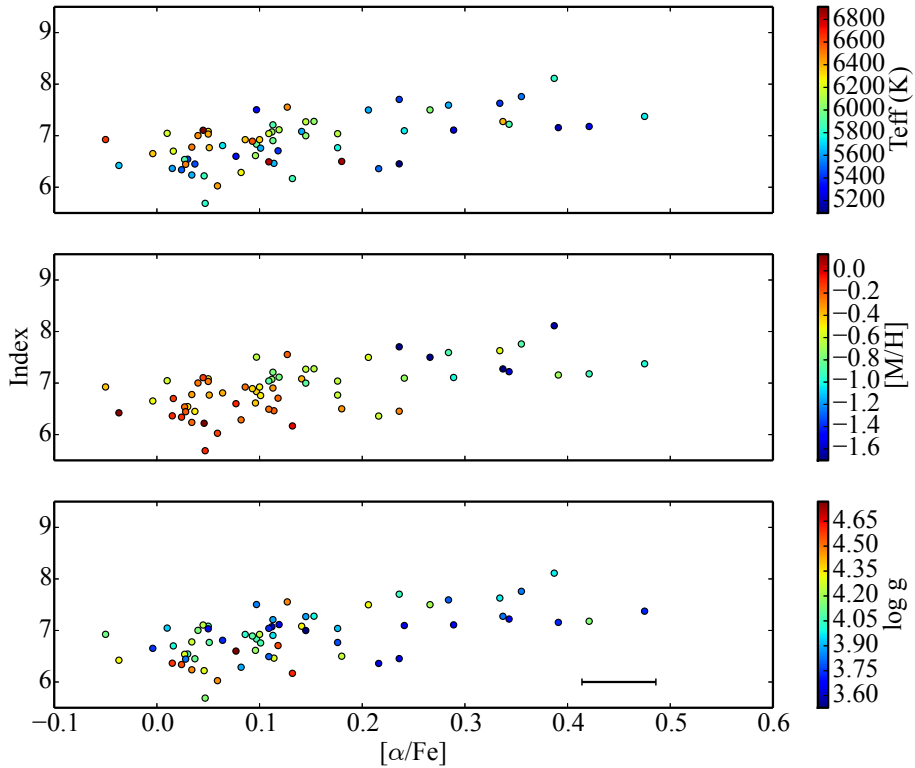}
\caption{The index value as a function of [$\alpha$/Fe] broken down by Teff (top panel), [Fe/H] (middle panel), log g (bottom panel) for 73 SDSS stars with parameters obtained by high resolution spectra that exist in our parameter range. The error bar represents the typical errors on the [$\alpha$/Fe] from the SDSS high-resolution validation set.}
\label{fig:SDSSindex}
\end{figure*}
We studied the index as a function of [$\alpha$/Fe] and the stellar parameters in Figure \ref{fig:SDSSindex}. The colour scheme and axes are the same as Figure \ref{fig:Tspecindex}. We see again that it is possible to separate stars based on their $\alpha$-abundances using the index alone as long as the star exists inside our parameter range. The index value needed to separate stars into $\alpha$-rich and $\alpha$-poor populations is consistent with the tests on the synthetic grid as well as the ELODIE and N10 validation sets. 
\subsection{Converting the Index to an Estimate of [$\alpha$/Fe]}
\label{subsec:convertindex}
We have shown, for all of our tests with synthetic and real spectra, that there is a linear relationship between our index and [$\alpha$/Fe]. We used a linear regression model to approximate the index as a function of [$\alpha$/Fe], to estimate the [$\alpha$/Fe] directly from the index. The ELODIE validation set was used as a calibration set because it is the largest one. A linear regression model for the ELODIE dataset yields:

\begin{equation}
\textrm{Index} = \mathrm{[\alpha/Fe]} \times 4.32 + 6.28
\label{eq:convert}
\end{equation}

The regression model achieved a correlation coefficient of 0.86 and is statistically significant with relatively low scatter (5 per cent in index space). The equation to convert our index into a direct estimate of [$\alpha$/Fe] was tested on the remaining validation sets (N10, and SDSS high-resolution targets). We also added an additional validation set using globular clusters and open clusters to compare the estimated [$\alpha$/Fe] with the high-resolution average measured in the literature. 

\begin{figure}
\includegraphics[width=\columnwidth]{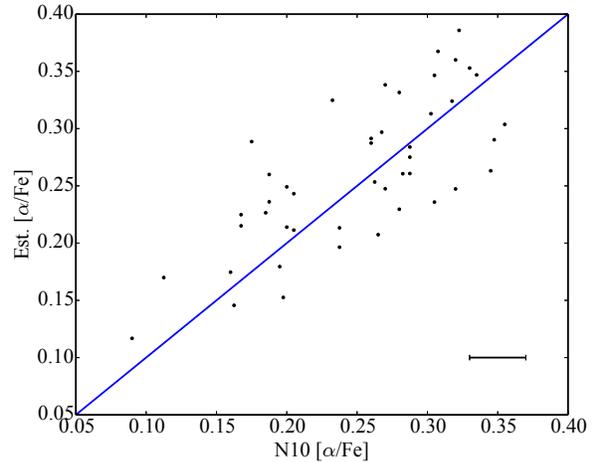}
\caption{The estimated [$\alpha$/Fe]  as a function of the high-resolution [$\alpha$/Fe] measurement from N10. The blue line represents a 1:1 relation. The error bar represents the typical errors on the [$\alpha$/Fe] from the N10 validation set.}
\label{fig:Nissenestalpha}
\end{figure}

\begin{figure}
\includegraphics[scale=0.35]{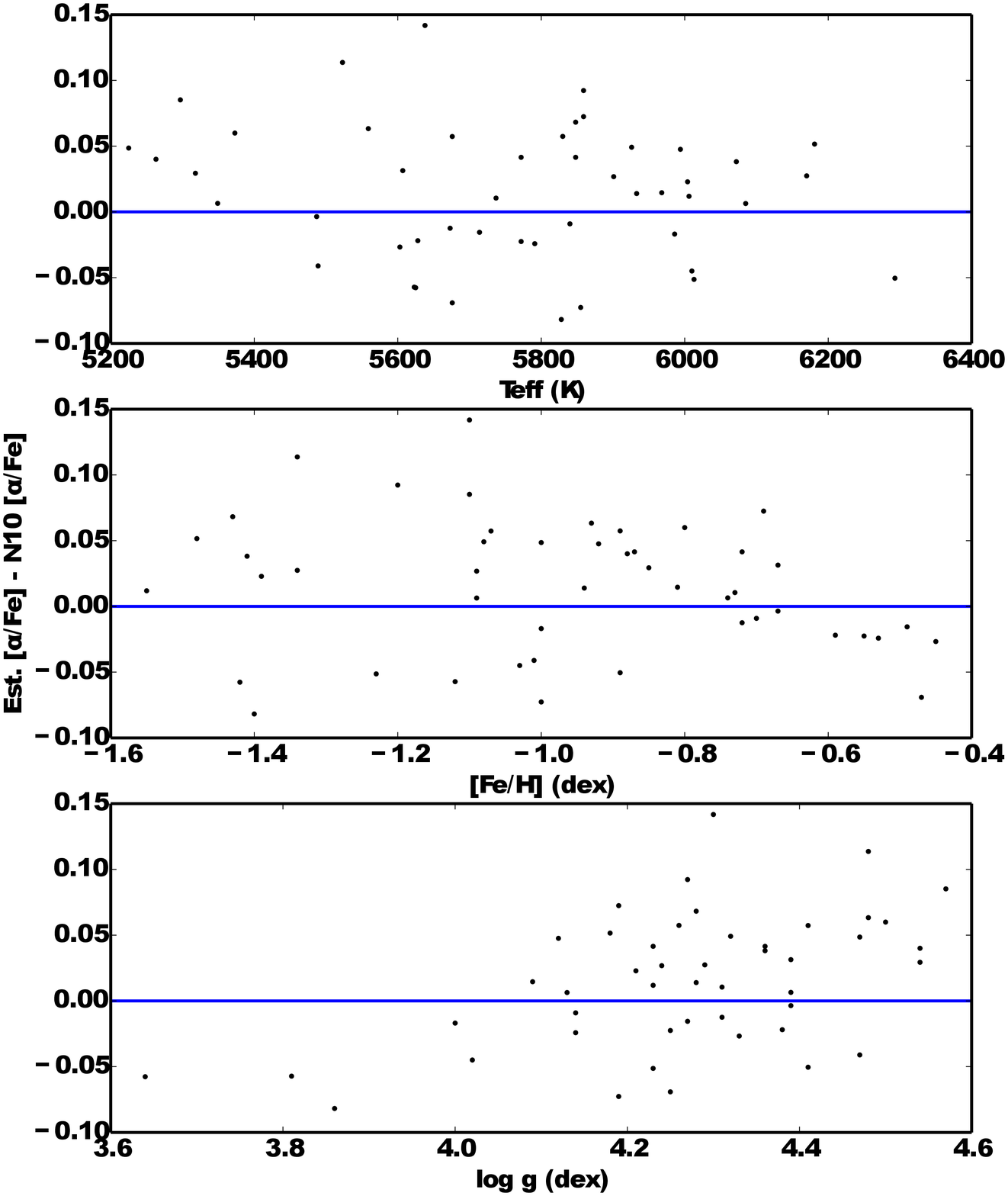}
\caption{The residual between the estimated and true $\alpha$/Fe] broken down by Teff (top panel), log g (middle panel), [Fe/H] (bottom panel) for the NOT/FIES subset of the N10 data.}
\label{fig:Nissensystematic}
\end{figure}

For the case of N10, we used the ELODIE dataset as a calibration set to convert our index in the [$\alpha$/Fe] through Equation 2, we found the [$\alpha$/Fe] and estimated [$\alpha$/Fe] are in good agreement with a mean offset of 0.01 dex and a external error of $\pm$0.05 dex (Figure \ref{fig:Nissenestalpha}). We also found no significant offsets or correlations between the residuals in [$\alpha$/Fe] and the stellar parameters (Figure \ref{fig:Nissensystematic}). Finally, we repeated the analysis for the SDSS calibration targets and found a very good agreement between our estimated $\alpha$-abundances and the high-resolution estimates with an offset in [$\alpha$/Fe]  = 0.00 dex and an external error of $\pm$0.10 dex. 

\subsection{Comparison with SDSS Clusters}
We used a sample of two globular clusters and one open cluster observed with the SDSS to validate our estimated [$\alpha$/Fe] values. Clusters are a good test bed to further validate our method, as each cluster should have an average [$\alpha$/Fe] with a relatively small spread. While clusters have small star-to-star variations in their abundances of light elements (C, N, Al, Na) we expect these to not strongly effect the $\alpha$-elements and thus the [$\alpha$/Fe] measurement. In order to achieve a large enough sample of stars in the clusters, we had to decrease the log g constraint to log g $>$ 3.0 dex. Further, we require the SNR to be larger than 40 (see Section \ref{subsec:SNReffect}) to only select out stars for which we can obtain suitable values of [$\alpha$/Fe]. 

The literature values for [$\alpha$/Fe] were complied by averaging the mean cluster abundances of the individual $\alpha$-elements for each cluster. The standard deviation was also computed and is displayed in Table \ref{tab:GCdata}. We estimated the [$\alpha$/Fe] for likely cluster members\footnote{For more information regarding the selection of likely cluster members in the clusters consult \cite{Lee2008b,Smolinski2011}} from M71 and M15, and the open cluster NGC 2420 using our spectral index method. The other clusters of \cite{Lee2011}, M13, M67, NGC 6791, are outside our metallicity range. For the three clusters that are in our parameter range, we estimated the mean and standard deviation in [$\alpha$/Fe]. Both the literature and estimated values for the mean [$\alpha$/Fe] can be found in Table \ref{tab:GCdata}. Our estimated mean [$\alpha$/Fe] for each cluster is in very good agreement with the literature values providing further proof that our  method can produce meaningful values for [$\alpha$/Fe]. 

\begin{table}
\caption{Globular/Open Cluster data.} 
\begin{tabular}{l c c c c}
\hline\hline
Cluster  & $<$[Fe/H]$>_{lit}$ & $<$ [$\alpha$/Fe]$>_{lit}$ & $<$[$\alpha$/Fe]$>_{est.}$ & N$_{mem}$ \\
 & (dex) & (dex) & (dex) & \\
\hline
M13&-1.58 $\pm$ 0.04 & 0.20 $\pm$ 0.04   & 0.19 $\pm$ 0.18    & 57    \\
M71 & -0.80 $\pm$ 0.02  & 0.23 $\pm$ 0.10 & 0.13 $\pm$ 0.05   & 5             \\
NGC 2420 & -0.05 $\pm$ 0.02 & 0.03 $\pm$ 0.09   & 0.08 $\pm$ 0.08   & 72     \\
\hline      
\end{tabular}
\\
NOTE - The data for each of the clusters were complied from the following sources: M13:  \cite{Sneden2004}, \cite{Cohen2005}; M71:  \cite{Boesgaard2005}; NGC2420: \cite{Pancino2010}. N$_{mem}$ is the total number of stars in the SDSS that were used to compute the average abundances. 
\label{tab:GCdata}
\end{table}

\subsection{Computation of Internal and External Error on the Index}
The total internal error on the index at any SNR can be estimated by completing noise injection experiments on both the SDSS data and synthetic spectra. We do the later in Section \ref{subsec:SNReffect} and find the internal error can be as high as 0.60 ($\sim$ 10 per cent) at SNR = 40. Propagating this to estimate the uncertainty in [$\alpha$/Fe] via Equation \ref{eq:convert} yields $\sigma_{internal}= \pm$ 0.13 dex. The external uncertainty in the [$\alpha$/Fe] is estimated by using the validation sets is shown to be as high as $\sigma_{external} = \pm$ 0.1 dex. Thus a conservative estimate of the total uncertainty in [$\alpha$/Fe] is the internal and external errors added in quadrature and is $\sigma_{\mathrm{[\alpha/Fe] }} = \pm$ 0.16 dex at a SNR = 40. The external and internal errors are comparable but the internal error can be greatly decreased by carefully selecting the range of stellar parameters. For example, when we constrained the temperature between 6000 and 6500 K and the metallicity between -0.8 and -2.0 dex, i.e. the expected turnoff region in our sample, we found that the index varies on the order of 0.20. This implies we can estimate [$\alpha$/Fe] with an internal error of $\pm$ 0.05 dex by constraining the stellar parameter to focus on the turnoff region only.

Our method improves on the current methods of \cite{Lee2011} and \cite{Franchini2011} by decreasing the dependence on models and uncertainties in the stellar parameters. We have shown that with our empirical spectral index method we can estimate [$\alpha$/Fe] to within $\sim \pm$ 0.1 dex when accounting for the full parameter space described in Section \ref{subsec:indexpreformance}, which has a temperature span of 2000 K, log g span of 1.5 dex and a metallicity coverage of 2.0 dex. In context, the spectral matching technique of \cite{Lee2011} has an internal error of $\sim \pm$ 0.1 dex in [$\alpha$/Fe] for a temperature uncertainty of only $\pm$ 300 K.

\section{Results}
\label{sec:Ages}
\subsection{Distribution of $\alpha$-elements in the Inner Halo}
\begin{figure}     
       \subfigure[]{
               \includegraphics[width=1\columnwidth]{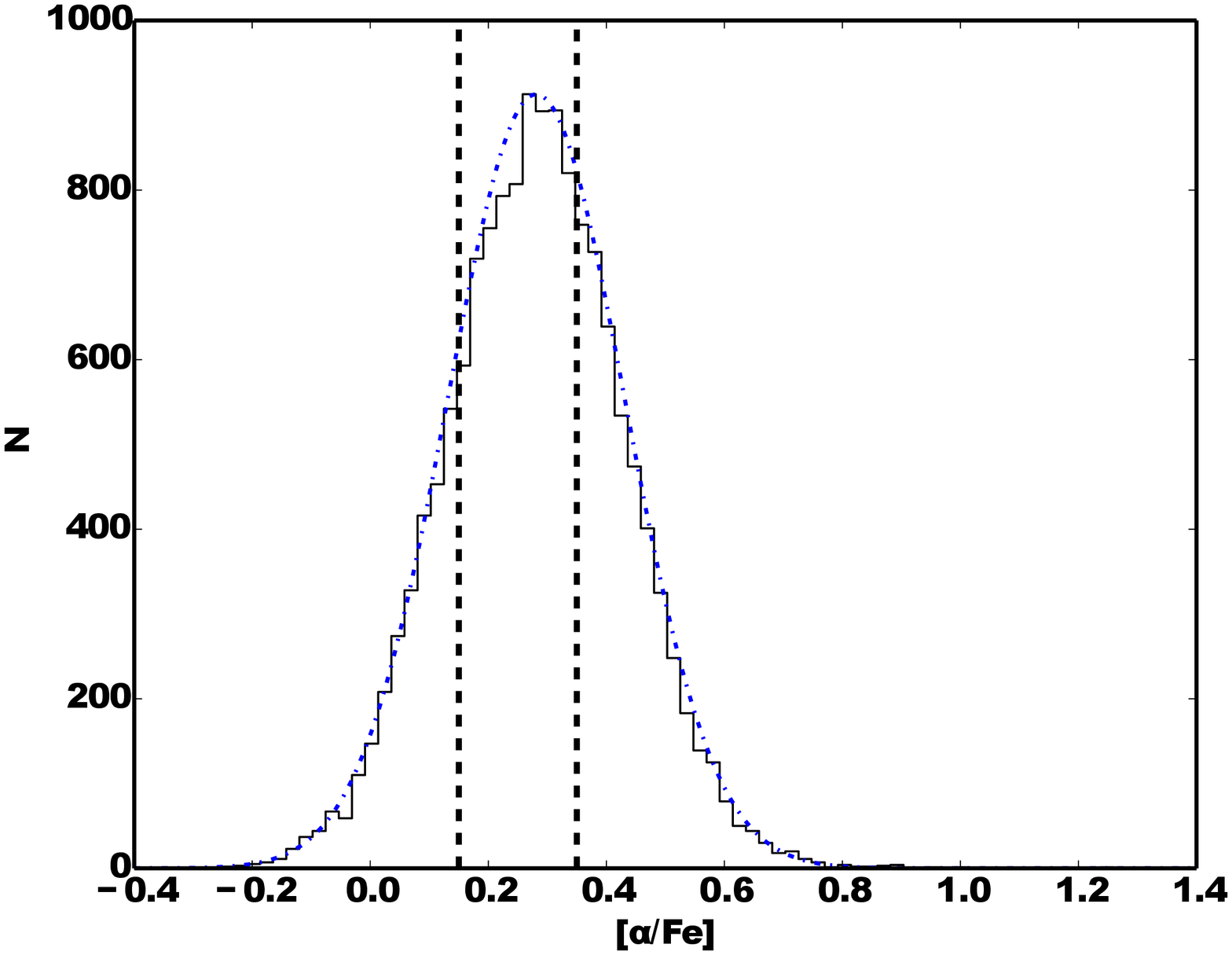}
                \label{fig:sampleadist}  }
        \subfigure[]{
                \includegraphics[scale=0.35]{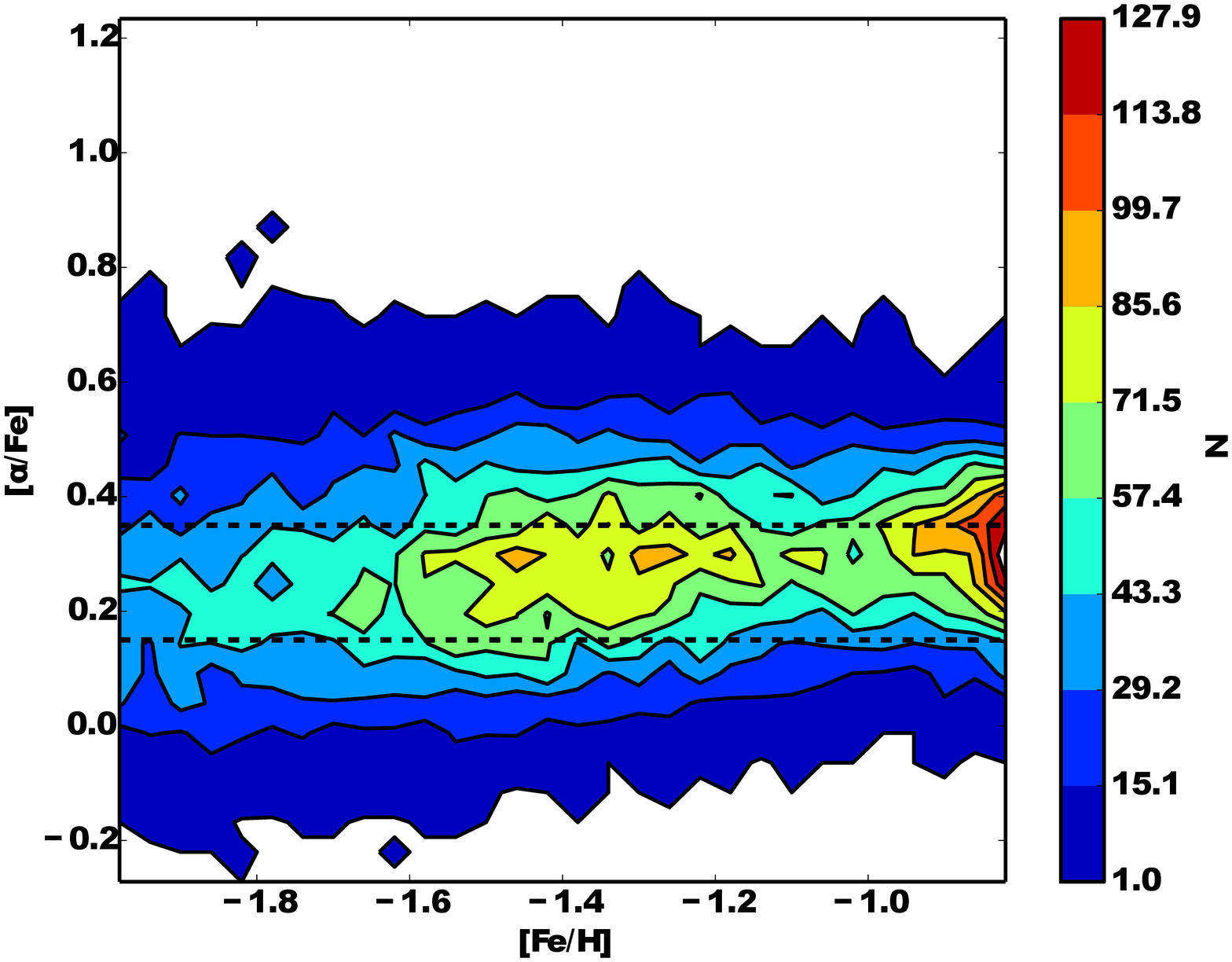}
               \label{fig:afeh}   }

        \caption{(a) The [$\alpha$/Fe] distribution of our sample of 14757 SDSS F and G stars. (b) The [$\alpha$/Fe] as a function of [Fe/H] for our sample. The [$\alpha$/Fe] was determined by converting the index via Equation \ref{eq:convert}. The dotted lines in both panels represent the cut in [$\alpha$/Fe] to obtain the $\alpha$-rich and $\alpha$-poor subsamples. The blue dash-dotted line in panel (a) is represents a single Gaussian fit to the distribution.}
        \label{fig:sampledists}
\end{figure}

The effective temperature and metallicity for each of our targets are sourced from the adopted values from the SDSS SSPP. We computed the [$\alpha$/Fe] for our targets using the spectral index method. The estimated [$\alpha$/Fe] distribution and metallicity-[$\alpha$/Fe] diagram are shown in Figure \ref{fig:sampleadist} and Figure \ref{fig:afeh}, respectively. The peak of the [$\alpha$/Fe] in Figure \ref{fig:sampleadist} is +0.30 dex which confirms that halo stars are, on average, $\alpha$-enhanced with a large sample from SDSS \citep[and references therein]{Helmi2008, Feltzing2013}. The dash-dotted line in Figure \ref{fig:sampleadist} shows the Gaussian fit to the [$\alpha$/Fe] distribution function. It is interesting to point out that the fit over predicts the amount of low-$\alpha$ stars near [$\alpha$/Fe] = 0.20 dex. We plan to explore this and the relative ratio of these two populations as a function of Galactic parameters further in the next work of this series. Constraining the sample to just the turnoff region, i.e. 6000 $<$ T$_{\mathrm{eff}}$ $<$ 6500 K, log g $>$ 3.5 dex, -0.8 $<$ [Fe/H] $<$ -2.0 dex, produced a standard deviation of the distribution in [$\alpha$/Fe] of 0.15 dex. This dispersion is larger than the estimated uncertainty in the constrained parameter space indicating we can resolve the $\alpha$-rich and $\alpha$-poor populations. 

Motivated by the conversion described in Section \ref{subsec:convertindex} and N10, $\alpha$-poor stars are defined as any stars which have an index less than 6.93, corresponding to an [$\alpha$/Fe] $<$  +0.15 dex and $\alpha$-rich stars as any stars which have an index larger than 7.79, corresponding to an [$\alpha$/Fe]  $>$ +0.35 dex. These criteria are designed to account for the boundary of $\alpha$-rich and $\alpha$-poor stars at $\sim$ [$\alpha$/Fe] = +0.25 dex (N10) and the uncertainty in the estimated [$\alpha$/Fe]. The criteria are used to select, in a statistical way, stars which are have a high probability of being either $\alpha$-rich or $\alpha$-poor based on the 1-$\sigma$ uncertainties in the estimated [$\alpha$/Fe].

\subsection{Turnoff Detection and its Uncertainties}
The turnoff temperature can be used to determine the age of the youngest stellar population \citep[for further discussion on this method consult][]{Soderblom2010}. We used a Sobel-Kernel edge detector algorithm to determine the temperature of the main sequence turnoff \citep[for more details consult][]{Jofre2011}. The method assumes the temperature distribution of a stellar population will display a very sharp decline (i.e. edge) near the turnoff as the more massive, hotter stars in the population have evolved off the main sequence. The edge, or turnoff temperature in our case, was determined by constructing a temperature distribution function for a range of metallicity bins and computing maximum of its derivative. This algorithm is sensitive to the sampling and in order to determine the uncertainty in the turnoff temperature we used a bootstrap method \citep{Jofre2011, Tabur2009}. In the bootstrap, we randomly re-sampled 80 per cent of our effective temperature distribution and recomputed the turnoff temperature via the Sobel-Kernel edge detector. We computed 500 iterations to find the variation in the turnoff temperature as a result of the sampling. Typical 3-$\sigma$ uncertainties in the turnoff temperature from the bootstrap are on the order of 60 K.

The number of stars sampled near the turnoff primarily affects the bootstrapping error. The more stars near the turnoff, the better sampled it is and the lower the overall bootstrapping error becomes. This can be accomplished by increasing the metallicity bin. However, since the turnoff temperature depends on metallicity, increasing the metallicity bin size can cause a noisier turnoff. \cite{Jofre2011} used a Monte Carlo approach and showed that the errors on the metallicity measurement will also induce a significant effect on the turnoff detection error because this method relies on binning in both temperature and metallicity space. We adopted this approach computing 500 realizations including the errors on the metallicity. The results of this simulation confirm that if the bin size is smaller than $\sim2\sigma_{[Fe/H]}$ then an individual star in our sample can jump between different metallicity bins and thus change the detected turnoff temperature significantly \citep{Jofre2011}. On the other hand, excessively large bin sizes can lead to a less precise turnoff temperature because the turnoff temperature is metallicity-dependent. Typical uncertainties in [Fe/H] at these higher SNR are on the order of 0.15-0.2 dex. Our simulations showed that the uncertainty in the turnoff temperature due to the metallicities becomes small relative to the bootstrapping error as the metallicity bin size becomes larger than $\sim$ 0.3 dex. Therefore, the uncertainty in the turnoff temperature is defined as the 3$\sigma$ error as a result of the bootstrapping with a bin size in metallicity of 0.3 dex. 

\subsection{Metallicity - Temperature Diagram}

\begin{figure*}
\includegraphics[scale=0.35]{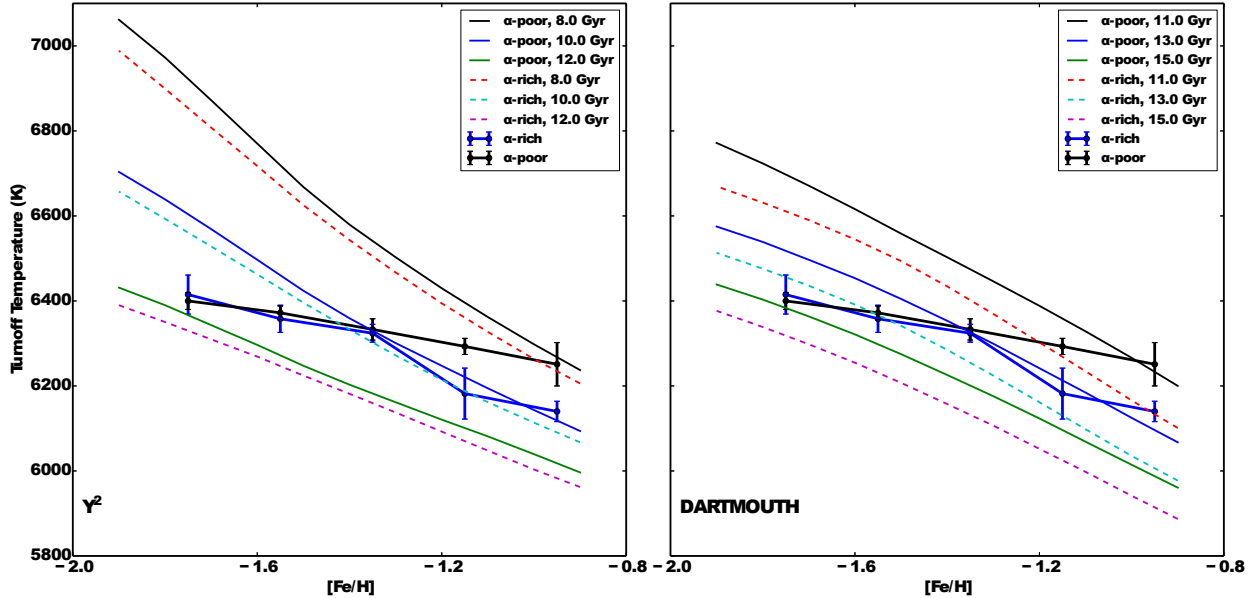}
\caption{Metallicity as a function of turnoff temperature withY$^2$ isochrones (left panel) and Dartmouth isochrones over plotted (right panel). }
\label{fig:mettemp}
\end{figure*}

In the left panel of Figure \ref{fig:mettemp}, we plot the turnoff temperature as a function of metallicity of our SDSS sample over plotted with the $\alpha$-rich and $\alpha$-poor Dartmouth \citep{Dotter2008} isochrones. The right panel of Figure \ref{fig:mettemp} shows the turnoff temperature as a function of metallicity for our sample over plotted with the Yonsei-Yale \citep[Y$^2$,][]{Demarque2004} isochrones. The first thing to note is that the turnoff temperature as a function of metallicity does not follow along an isochrone of a constant age indicating there is a correlation between age and metallicity. These plots also show that the turnoff temperature of the $\alpha$-rich and $\alpha$-poor stars in our sample are, within the errors, equal at low-metallicity. As the metallicity increases, upwards of $\sim$ -1.4 dex the $\alpha$-rich stars have a significantly lower turnoff temperature indicating they are older than the $\alpha$-poor stars. All of our stars are older 8 Gyr with both isochrones. The Dartmouth isochrones produce larger ages compared to the Y$^2$ isochrones and in some cases have ages large than the accepted age of the universe \citep[13.8 Gyr,][]{planck2013}. This is likely a result of the prescription of atomic diffusion \citep{Jofre2011}.

Qualitatively, the turnoff temperature of the $\alpha$-rich component as function of metallicity is not significantly different with the results of \cite{Jofre2011}. This is expected as the bulk of the stars in the sample are $\alpha$-rich. Another interesting finding we observe from the over plotted isochrones is that the Dartmouth isochrones have larger differences in the turnoff temperature between $\alpha$-rich and $\alpha$-poor stars at the same age. This indicates that the two isochrones have different responses to $\alpha$-enhancement. The different isochrone sets also use different helium mixtures, which can also affect the isochrones.

\subsection{Isochrone Analysis: Ages and their Errors} 
With the turnoff temperatures, metallicities and a set of isochrones in hand, we can determine the age of our $\alpha$-rich and $\alpha$-poor populations. For this we used the Y$^2$ isochrones \citep{Demarque2004} which were interpolated to determine the theoretical turnoff temperature as a function of age, metallicity, and [$\alpha$/Fe]. The Y$^2$ isochrones include atomic diffusion \citep{Demarque2004} while the Dartmouth isochrones only include moderate atomic diffusion \citep{Dotter2008}. As a result, the Dartmouth isochrones produce larger ages compared to the Y$^2$ isochrones. To avoid dependencies in the isochrones used, we have considered the relative ages between the $\alpha$-rich and $\alpha$-poor populations.  

The effect of increasing [$\alpha$/Fe] in an isochrone model, while keeping [Fe/H] fixed, is to effectively increase the total metal abundance. Dartmouth and Y$^2$ isochrones show that increasing the $\alpha$-abundance will cause the turnoff temperature to decrease at a constant age. Physically, this can be attributed to the increase in the importance of the CNO cycle in energy generation causing an earlier hydrogen exhaustion for $\alpha$-enhanced stars \citep{Kim2002}. 

The ages for our $\alpha$-rich and $\alpha$-poor subsamples were determined by combining the turnoff temperature and the Y$^2$ isochrones. We choose the Y$^2$ isochrones because they include more physics (diffusion, etc.) than the corresponding Dartmouth isochrones. The metallicity of the isochrones used was equal to the metallicity of the bin centre. The [$\alpha$/Fe] was set to +0.40 dex and +0.20 dex for the $\alpha$-rich and $\alpha$-poor subsamples, respectively. The uncertainty in the age was determined by propagating the uncertainty in the turnoff temperature. In this case, the uncertainty in the turnoff temperature was determined by the bootstrap and Monte Carlo analysis in metallicity. We also considered the uncertainty in age due to metallicity by adding in quadrature the uncertainty in turnoff temperature as a result of metallicity errors. The final turnoff temperature, error in the turnoff temperature, ages, and uncertainty in the ages for our $\alpha$-rich and $\alpha$-poor subsamples are shown in Table \ref{tab:age+TO}. For our analysis, we used normalized ages, which we define as the absolute age divided by the mean age of all of the stars (10.07 Gyr), because of the large theoretical uncertainties in the ages determined from the turnoff temperature. Our goal is to quantify the age difference so using normalized ages is appropriate. It is important to point out that the error on the age we quote is only the internal error.

\begin{table}
\caption{Turnoff Temperature and Ages for our SDSS FG dwarf sample using Y$^2$ models.} 
\begin{tabular}{c c c c c c }
\hline\hline
[Fe/H] & $T_{\mathrm{eff,TO}}$  & $\sigma_{\mathrm{T_{eff,TO}}}$ & Age & $\sigma_{\mathrm{Age}} $& Normalized Age \\
(dex) & (K) & (K)&(Gyr) & (Gyr) & \\
\hline 
&  & $\alpha$-rich & & & \\
\hline \hline
-0.95 & 6140 & 24 & 9.18 & 0.27&0.91\\
-1.15 & 6182 & 60 &10.09 & 0.64&1.00\\
-1.35 & 6324 & 21 & 9.73 & 0.15&0.96\\
-1.55 & 6358 & 32 & 10.71 & 0.27&1.06\\
-1.75 & 6415 & 46 & 11.20 & 0.25&1.11\\

\hline
&  & $\alpha$-poor& &  &\\
\hline \hline
-0.95 & 6251 & 51 & 8.23 & 0.32&0.82\\
-1.15 & 6293 & 19 & 9.04 & 0.13&0.90\\
-1.35 & 6333 & 25 & 9.98 & 0.20&0.99\\
-1.55 & 6372 & 16 & 10.89 & 0.12&1.08\\
-1.75 & 6400 & 20 & 11.67 & 0.11&1.16\\

\hline
\hline      
\label{tab:age+TO}
\end{tabular}
\\
\label{tab:TO+Age}
\end{table}

\section{Discussion}
\label{sec:discussion}
\subsection{Age-Metallicity Relation}
\label{subsec:Agemetal}
\begin{figure}
	\includegraphics[width=1\columnwidth]{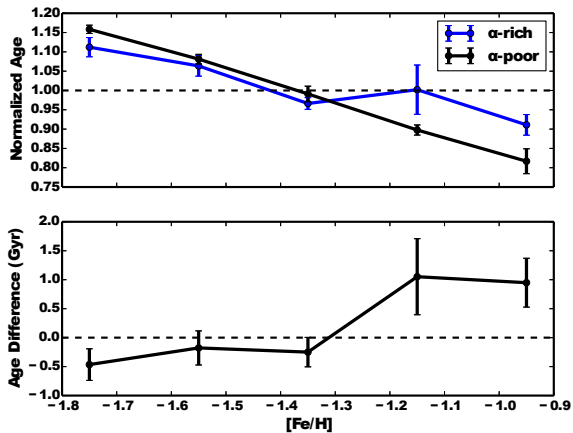}
	\caption{Top panel: normalized Y$^2$ isochrone ages as a function of metallicity for $\alpha$-rich (blue) and $\alpha$-poor (black) populations. Bottom Panel: age difference between the $\alpha$-rich and $\alpha$-poor populations as a function of metallicity.}
	\label{fig:agemetal}
\end{figure}

We have, for the first time, measured the age-metallicity relation of an $\alpha$-rich and $\alpha$-poor population in the Galactic halo at metallicities between -0.80 dex and -2.0 dex (Figure \ref{fig:agemetal}). We note that: (1) at high metallicities ([Fe/H] $>$ -1.4 dex), the $\alpha$-rich population is older than the $\alpha$-poor population, and (2) at low metallicities ([Fe/H] $<$ -1.4 dex), the two populations are coeval within the errors. We have found that the shape of the age-metallicity relation differs for both $\alpha$-populations. The $\alpha$-poor stars tend to have a steeper correlation between age and metallicity compared to the $\alpha$-rich stars. This is consistent with the idea that the $\alpha$-poor stars were formed in areas with slower chemical evolution compared to their $\alpha$-rich counterparts. The very shallow slope on the age-metallicity correlation for $\alpha$-rich stars seems to infer they were formed in a quick event with high star formation rates which were able to produce a broad range in metallicities in a short time (less than $\sim$ 10$^9$ years) to keep the $\alpha$-abundance enhanced. By testing multiple isochrone sets, we have found that our age sequence at low metallicities, namely that the $\alpha$-rich population is coeval with the $\alpha$-poor population, within the errors, is robust. Our age difference between the $\alpha$-rich and $\alpha$-poor populations in the most metal-rich bin is affected by as much as 1 Gyr. However, this difference does not effect the interpretation of our results. 

It is also worth commenting that the age-metallicity relation that we observe in Figure \ref{fig:agemetal}, is similar to the 'Y'-like bifurcation seen in the globular clusters of \cite{Marin-Franch2009} and \cite{Leaman2013}.  \cite{Marin-Franch2009} found two well defined tracks in the age-metallicity diagram for globular clusters: a 'young' track which can be associated with dwarf galaxies and an 'older' track with a small age dispersion which can be associated with an in situ population that may have formed from a protogalactic gas cloud. We found a similar bifurcation in our halo field star sample which supports a scenario of common origin for the GC and halo field stars \citep[e.g.][]{Martell2010}. 

\subsection{Implications for the Formation of the Galactic Halo}
One unambiguous result from our study is that all of the stars in our sample, regardless of which isochrone set is used, are older than $\sim$ 8 Gyr. This confirms the notion that the Galactic halo was formed and assembled very early on as suggested by both theory \citep{Robertson2005, Font2006a, Zolotov2009, Zolotov2010, Font2011} and observations \citep[e.g.][]{Schuster2012}.

The theoretical work of \cite{Robertson2005} and \cite{Font2006a} lays out what \cite{Schuster2012} refer to as a 'dual-accretion' scenario. In this scenario, the halo is built from the bottom-up via hierarchical merging. The dual-accretion models produces a Galactic halo by merging smaller sub-units of stars together rapidly at early times fully destroying the sub-units in the process. The proto-halo systems form massive stars rapidly and are enriched primarily by Type II supernova producing an $\alpha$-rich population. Since the sub-units are broken up quickly, there is little iron enrichment via Type Ia supernova leaving behind a set of low metallicity, $\alpha$-rich stars. These would correspond to the $\alpha$-rich stars in our sample. Longer-lived dwarf galaxies, on the other hand, are polluted via both types of supernova leading to a population of stars that have a wide-range in metallicities and are $\alpha$-poor. These stars are accreted onto the Milky Way at later times and could represent the $\alpha$-poor stars in our sample. 

Interestingly, the age-metallicity diagram from the theoretical work of \cite{Font2006a} appears to have two tracks, one that has a steeper slope in age-metallicity space than the other which is the case in our sample (Figure \ref{fig:agemetal}). This bifurcation in the age-metallicity relation can be interpreted as evidence that the Galactic halo was built in two phases. One quick phase that produced the $\alpha$-rich stars over broad metallicity ranges, leading to a shallow age-metallicity relation. Followed by a second phase of accretion of stars that evolved in environments of slower chemical evolution, leading to a more pronounced age-metallicity relation. 

The more recent hydrodynamic simulations by \cite{Font2011} have predicted that the accreted stars are, on average, 3-4 Gyr older than the in situ population. If we make the assumption that the in situ population are tracked by our $\alpha$-rich stars \citep[an assumption often made in the literature, e.g.][]{Schuster2012} than our relative age sequence would indicate that the accreted stars are $\sim$ 1 Gyr \textit{younger} than the in situ population. This result does not agree with the results of \cite{Font2011} and is only about half 2-3 Gyr age difference observed by \cite{Schuster2012}. However, this discrepancy could be remedied if we do not make the assumption that $\alpha$-rich stars are the in situ population and the $\alpha$-poor stars are the accreted population or if the feedback in their simulations were less efficient in low-mass halos \citep{Font2011}. The unavoidable (larger) errors on [$\alpha$/Fe] with low-resolution spectra can lead to a slight mixing of the two populations which could also explain the diminished age difference compared to the results of \cite{Schuster2012}.

Contrary to \cite{Robertson2005} and \cite{Font2006a, Font2011}, the models of \cite{Zolotov2009, Zolotov2010} predicted that the in situ population within the Galactic halo is likely formed within inner 4 kpc of the galaxy's centre and is formed via cold, in-falling gas. The in-falling gas creates a primeval bulge or disk. After which those stars in the primeval disk are kinematically heated into the halo via early accretion and merging events which at the same time populate the accreted component of the halo via tidal stripping \citep{Purcell2010}. These in situ stars can be interpreted as our $\alpha$-rich stars \citep[just as in][]{Schuster2012}. Sars formed in the dwarf galaxies that later merged with the primeval disk/bulge can be attributed to our $\alpha$-poor stars. \cite{Zolotov2010} found, in all but one of their model halos, the in situ population was formed early on and the accreted population was formed, on average, at later times and then subsequently accreted onto the halo. However, it is not clear whether this is true at all metallicities or just higher metallicities. Our relative age sequence at high metallicities, namely $\alpha$-rich stars being older than $\alpha$-poor stars on average, favors the models of \cite{Zolotov2009, Zolotov2010} and at high metallicities confirms the observational results of \cite{Schuster2012} with a larger, statistical sample. Further the bifurcation we observed should still be present in these models. The results of this study can be used to help construct more realistic models and better constrain the basic physics, in particular feedback mechanisms, which can replicate the observations. 

\section{Summary}
\label{sec:summary}
Motivated by the results of both \cite{Schuster2012} and \cite{Jofre2011}, we addressed the intriguing problem of the relative ages of the $\alpha$-rich and $\alpha$-poor stars in the Galactic halo with a large sample of stars from SDSS. To that end, we developed a new spectral index-based method to estimate the [$\alpha$/Fe] abundances using low resolution (R $\sim$ 2000), moderate SNR ($>$  40) SDSS spectra. We studied the ages of a sample of main sequence turnoff halo field stars selected using a cuts in colour and Galactic latitude. With our method (described in Section \ref{sec:Method}), we could split the halo field star population statistically into an $\alpha$-rich and $\alpha$-poor subsample. We used a Sobel-Kernel edge detection method to determine the turnoff temperature, and thus ages of the two stellar subsamples. Our results can be summarized in the following points:
\begin{enumerate}
\item A spectral-index based method (see Section \ref{sec:Method} for more details) was constructed  to rank stars based on [$\alpha$/Fe]. We have estimated the uncertainties of the [$\alpha$/Fe] to be on the order of $\sim \pm$~0.15 dex. This method can be used on a range of stellar parameters and is a semi-empirical method to estimate [$\alpha$/Fe] that automatically accounts for the stellar parameters. The uncertainty in [$\alpha$/Fe] from our method is comparable to the other methods \citep[e.g.][]{Lee2011}. Using this method, we found that the halo is comprised of an $\alpha$-rich and $\alpha$-poor population which may peak near [$\alpha$/Fe] $\sim$ +0.40 dex and +0.20 dex respectively (see Figure \ref{fig:sampledists}). It may be possible to extend this basic method to find stars with spectral peculiarities (e.g. s-process enriched star, etc.) which is something we plan to explore in future. Finally, this method can be expanded to other elemental species and spectral-types. 

\item Given the large absolute age, based on the Y$^2$ isochrones which included $\alpha$-enrichment and atomic diffusion, the Galactic halo must have been formed very early on. It is also likely that the Galaxy is a relatively quiet place not having undergone a major merging event in the last 8 Gyr. This is consistent with other observations and theoretical studies \citep[e.g.][]{Schuster2012, Robertson2005, Font2006a, Zolotov2009, Zolotov2010}.

\item We have made a first measure of the age-metallicity relation of halo field stars separated by $\alpha$-abundances at low metallicities. There appears to be a difference, on the order of 1 Gyr, in the ages of the $\alpha$-rich and $\alpha$-poor subsamples (see Section \ref{sec:Ages}). Using the Y$^2$ isochrones we found the $\alpha$-rich subsample is older than the $\alpha$-poor subsample in the high-metallicity case ([Fe/H] $\gtrsim$ -1.4 dex) which confirms the observational results of \cite{Schuster2012} and the theoretical results of \cite{Zolotov2009, Zolotov2010}. However, we extended the results of \cite{Schuster2012} to lower metallicities and found there is a break around [Fe/H] $\sim$ -1.4 dex, and the $\alpha$-rich subsample becomes coeval with the $\alpha$-poor subsample. Interestingly, this bifurcation in the age-metallicity diagram is also seen in globular clusters. This hints that the $\alpha$-rich population, with a shallow age-metallicity relation, was formed in a rapid event with high star formation (e.g. collapse of a protogalactic gas cloud), while the $\alpha$-poor stars were formed in an environment with a slower chemical evolution timescale.  Our results also confirm the idea that the $\alpha$-rich subsample may be a population formed in situ while the $\alpha$-poor subsample may have formed in satellite galaxies and accreted into the Milky Way halo.

\end{enumerate}

The relative age sequence we found support the models of \cite{Zolotov2009,Zolotov2010}. These models point out that we can use the fraction of in situ and accreted stars to disentangle the importance of accretion events in the assembly of the Galactic halo. With this new spectral-index based method to estimate [$\alpha$/Fe] from low-resolution spectra, we are in a good position to study the relative fraction of $\alpha$-rich and $\alpha$-poor stars in the Galactic halo.  

With upcoming large-scale surveys such as Gaia \citep{Perryman2001}, Gaia-ESO \citep{Gilmore2012}, 4MOST \citep{de_jong2012}, and GALAH \citep{Anguiano2014}, and it will be possible to resolve large sample of stars in position, velocity, distance and chemical phase-spaces. The next generation of data will undoubtedly open new windows into studying the ages of individual halo field stars as a function of chemical abundances providing detailed insights of the early stages of our home galaxy.  

\bibliography{SDSSbib}

\section*{Acknowledgements}
We thank the anonymous referee that provided us with helpful comments.  We would like to thank P. Nissen for helpful comments and for the validation dataset he graciously provided us. K.H. is funded by the British Marshall Scholarship program and the King's College, Cambridge Studentship. This work was partly supported by the European Union FP7 programme through ERC grant number 320360. This study was made possible by the Sloan Digital Sky Survey.
\pagebreak
 
\label{lastpage}

\end{document}